\documentclass[aip, preprint,graphicx]{revtex4-1}
\usepackage{graphicx}
\usepackage{dcolumn}
\usepackage{bm}

\usepackage[utf8]{inputenc}
\usepackage[T1]{fontenc}
\usepackage{mathptmx}
\usepackage{etoolbox}

\usepackage{amssymb}
\usepackage{mathtools}
\usepackage{algorithmic}

\usepackage{textcomp}
\usepackage{xcolor}
\usepackage{subcaption}
\usepackage{chemformula} 
\usepackage{booktabs} 
\usepackage{bm}
\usepackage{multirow}
\usepackage{float}
\usepackage{hyperref}
\usepackage{array}
\usepackage{gensymb}
\usepackage{layouts}

\usepackage{amsmath}

\begin{document}

\title{Image-Force Barrier Lowering of Schottky Barriers in Two-Dimensional Materials as a Function of Metal Contact Angle} 

\author{Sarah R. Evans}
\affiliation{Department of Electrical and Computer Engineering, The University of Texas at Dallas, 800 W Campbell Rd., Richardson, Texas 75080, USA.}%
\affiliation{Department of Materials Science and Engineering, The University of Texas at Dallas, 800 W Campbell Rd., Richardson, Texas 75080, USA.}%
\author{Emeric Deylgat}
\affiliation{Department of Materials Science and Engineering, The University of Texas at Dallas, 800 W Campbell Rd., Richardson, Texas 75080, USA.}%
\affiliation{Department of Physics, KU Leuven, Kasteelpark Arenberg 10, 3001 Leuven, Belgium.}%
\affiliation{Imec, Kapeldreef 75, 3001 Heverlee, Belgium.}%
\author{Edward Chen}
\affiliation{Corporate Research, Taiwan Semiconductor Manufacturing Company Ltd., 168, Park Ave. II, Hsinchu Science Park, Hsinchu 300-75, Taiwan}
\author{William~G.~Vandenberghe}
\affiliation{Department of Materials Science and Engineering, The University of Texas at Dallas, 800 W Campbell Rd., Richardson, Texas 75080, USA.}%
\email[]{william.vandenberghe@utdallas.edu}

\date{\today}

\begin{abstract}
Two-dimensional (2D) semiconductors are a promising solution for the miniaturization of electronic devices and for the exploration of novel physics. However, practical applications and demonstrations of physical phenomena are hindered by high Schottky barriers at the contacts to 2D semiconductors. While the process of image-force barrier lowering (IFBL) can considerably decrease the Schottky barrier, IFBL is not fully understood for the majority of prevalent contact geometries. We introduce a novel technique to determine the IFBL potential energy with application spanning far beyond that of any existing method. We do so by solving Poisson’s equation with the boundary conditions of two metal surfaces separated by an angle $\Omega$. We then prove that our result can also be obtained with the method of images provided a non-Euclidean, cone-manifold space is used. The resulting IFBL is used to calculate the expected contact resistance of the most prevalent geometric contacts. Finally, we investigate contact resistance and show how the stronger IFBL counteracts the effect of larger depletion width with increasing contact angle. We find that top contacts experience lower contact resistance than edge contacts. Remarkably, our results identify tunable parameters for reducing Schottky barriers and likewise contact resistance to edge-contacted 2D materials, enhancing potential applications.

\end{abstract}


\maketitle 
 
\section{Introduction}
\begin{figure}[t]
    \centering
    \includegraphics[width=8.6cm]{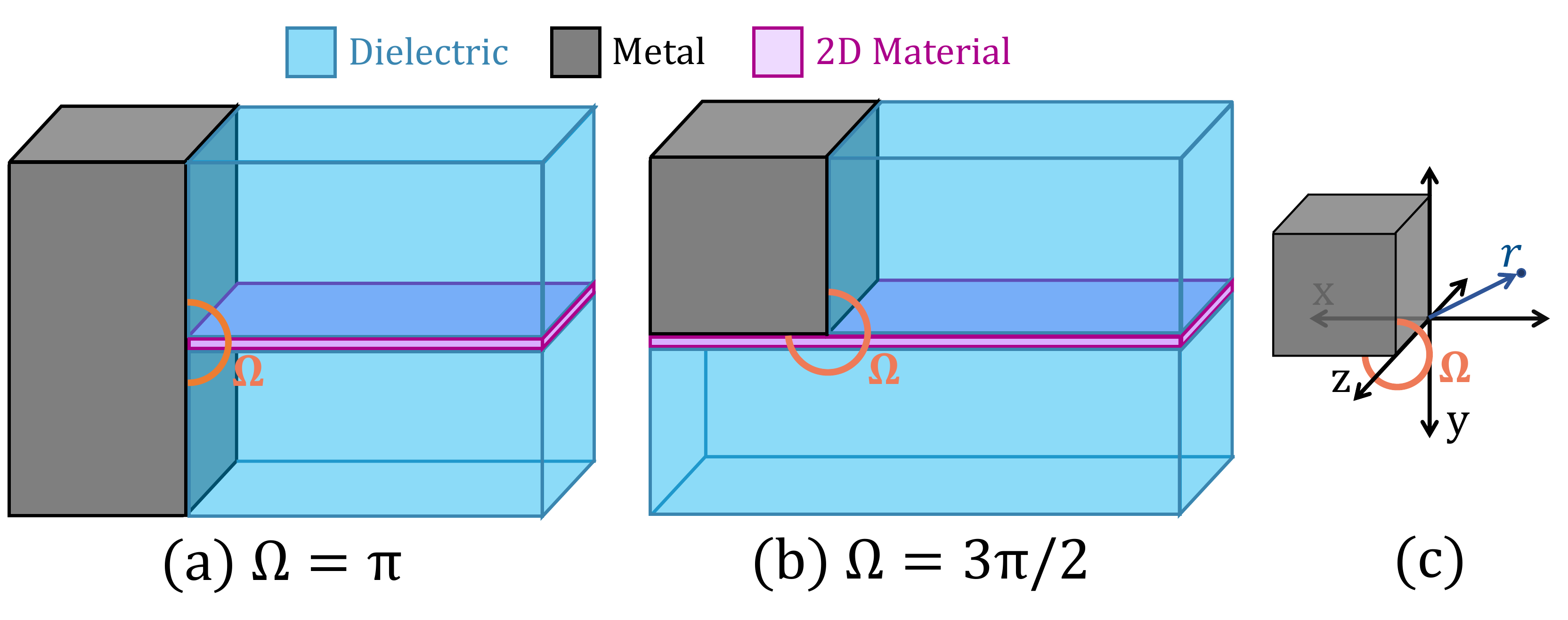}
    \caption{Illustration of a metal making (a) contact with the edge of a 2D material and (b) contact with the top of a 2D material. (c) shows the coordinate system we use in the derivation indicating Cartesian and cylindrical coordinates as well as the angle between the two metal surfaces, $\Omega$.}
    \label{fig:TopAndSideContacts}
\end{figure}

Two-dimensional (2D) semiconducting materials have garnered significant interest in the past couple of decades due to their unique properties.\cite{fiori2014electronics} For instance, graphene is a 2D Dirac semi-metal with exceptional mobility,\cite{yankowitz2019van, gosling2021universal, wu2008epitaxia1} 2D topological insulators have been realized showing quantized conductance at room temperature,\cite{vandenberghe2017imperfect, hasan2010colloquium} and Transition-Metal Dichalcogenides (TMDs) have unique optical properties and are being explored as the channel material for transistors.\cite{koperski2017optical, huang20222d, andrews2020improved, su2021layered, kang2014computational} Despite these promising characteristics, the high Schottky barriers experienced by 2D material contacts inhibit practical device applications and demonstrations of novel physics. 

Improving the Schottky barrier height to 2D material contacts has the potential to enhance several applications of 2D materials, including boosting quantum yields of qubits,\cite{lee2021boosting} enhancing optoelectrical coupling in 2D materials,\cite{lopez2013ultrasensitive} and improving heterostructures based on 2D materials.\cite{ang2018universal} Additionally, lowering the Schottky barrier is crucial for reducing the contact resistance experienced at 2D contacts.\cite{banerjee2020reducing, popov2012designing} Lower contact resistance can improve nanotube performance,\cite{hafizi2017band} aid connections of high-mobility graphene structures through metallic contacts,\cite{cusati2017electrical} and improve 2D electrical device performance.\cite{allain2015electrical, fiori2014electronics, shen2021ultralow} Therefore, it is paramount for the future of 2D device applications that methods which can lower the Schottky barrier, and likewise contact resistance, are discovered and fully understood.

A significant amount of attention has recently been paid to the modeling and understanding of the Schottky barrier seen in 2D materials.\cite{ang2018universal, venica2018adequacy, park2018review, trushin2018theoery, tongay2012Rectification, parto2021one, van2015Padilha} A prevalent approach to modeling the Schottky barrier is to use first principles density-functional theory (DFT) quantum transport simulations.\cite{van2015Padilha, courtin2020origin, er2020atomistic} However, DFT is limited in its ability to provide a complete and accurate Schottky barrier height as it does not include the effects of image force barrier lowering (IFBL) unless transport calculations are performed at the GW level.\cite{li2009gw}

Figure~\ref{fig:TopAndSideContacts}a-b shows two configurations for a metal to contact a 2D material surrounded by a dielectric, where Fig.~\ref{fig:TopAndSideContacts}b is more commonly used experimentally.\cite{chung2019experimentally} To date, the structure in Fig.~\ref{fig:TopAndSideContacts}a has the only known expression for the IFBL potential energy: $U_{\rm IFBL}^{\rm (a)}=-e^2/(16\pi \epsilon r)$,\cite{sze2021physics} where $e$ is the electron charge and $\epsilon$ the permittivity. The IFBL expression for Fig.~\ref{fig:TopAndSideContacts}b, however, is not available.\cite{vaknin2020schottky, yang2019horizontal}

In this work, we derive a general expression for the IFBL potential energy of a metal whose surfaces are separated by an angle $\Omega$ that can be applied to Fig.~\ref{fig:TopAndSideContacts}b. We do so using two different approaches: first by solving the Poisson equation in the presence of metal surfaces, and second by applying the method of images to a metal surface in a cone manifold. We solve Poisson's equation using the Kontorovich–Lebedev transform and calculate by how much the IFBL potential energy is weakened/strengthened compared to the known expression $-e^2/(16\pi \epsilon r)$.\cite{kontorovich1938one} We find that the IFBL is reduced by a factor $6-2/\sqrt{3}\approx0.85$ along the $x$-direction when using a contact as shown in Fig.~\ref{fig:TopAndSideContacts}b compared to the contact in Fig.~\ref{fig:TopAndSideContacts}a. We also find that the IFBL can be significantly improved when using a metal with a smaller $\Omega$, pointing to the potential of engineering the angle of the metal to lower the Schottky barrier. Further, because 
IFBL is scaled by the permittivity of the surrounding dielectric material, our results highlight the importance of choosing a dielectric with a low permittivity when contacting 2D materials. Finally, we calculate the contact resistance for contact geometries with differing angles. We show that the IFBL improvement is able to counteract worsening electrostatics as a function. For edge contacts, a non-perpendicular metal yields lower contact resistance, but top contacts experience lower contact resistance across all angles.

\section{Background}
IFBL is known to be a key mechanism in reducing the Schottky barrier height in contacts.\cite{sze1964photoelectric,binnig1984electron} The concept of IFBL originates from Schottky's proposition that electrons are attracted to the ``Thomson's image force'' emitted by a metal. Bethe then showed that IFBL also applied to metal-semiconductor contacts.\cite{schottky1914einfluss,bethe1942theory,rideout1978review} Subsequently, the IFBL effect has been confirmed experimentally in various metal-semiconductor or metal-insulator contacts.\cite{vaknin2020schottky, sze1964photoelectric} Recently, we demonstrated that the permittivity of the surrounding material can enhance the Schottky barrier lowering of 2D contacts, and using a low-permittivity dielectric can lower the Schottky barrier up to 50 times.\cite{brahma2022contacts, deylgat2022image} Therefore, IFBL plays a critical role in the Schottky barrier's height of 2D material contacts because IFBL is scaled by the permittivity of the surrounding material.

Historically, the method of images is used to calculate the IFBL potential energy. W. Thomson devised the method of images and showed that when an electron is near a metal and attracts a positive surface charge, the potential can be described by the sum of the Coulomb potential of the original particle and the Coulomb potential of an image located inside of the metal.\cite{thomson1848geometrical, maxwell1873treatise} The potential including the imaginary image exactly describes the classical potential profile outside of the metal. The method of images can be applied to metals that present symmetrically to the electrons, {\it i.e.} flat metal sheets and ellipsoidal-shaped metals. However, for an arbitrarily shaped metal, there is no general method available. 

An alternative to the method of images is to solve the Poisson equation with boundary conditions accounting for the metal.\cite{griffiths2005introduction,jackson1999classical} For a charge located at a single position, the Poisson equation reduces to the Laplace equation everywhere except at the location of the charge. In 2D, the vanishing Laplacian of holomorphic functions on the complex plane can be exploited and many solutions can be found using conformal mapping. However, for IFBL potential energy, the 3D Poisson equation needs to be solved and only separation of variables or a numerical approach can be used.

\section{Determining the Image Potential}
\subsection{Solving Poisson's Equation}
\begin{figure*}
\centering
    \includegraphics[width=17.2cm]{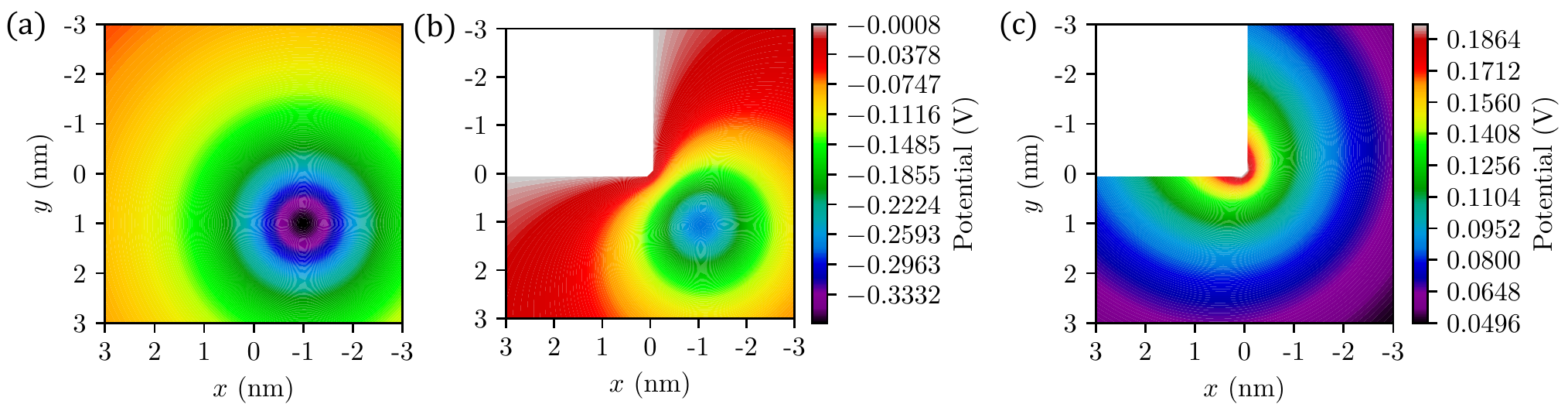}
    \caption{Graphs taken at $z=0$ of (a) $V_{\rm C}$, the potential caused by a single electron located at $(r_0,\theta_0,z_0)=(\sqrt{2},3\pi/4,1)$, (b) $V$, the potential an electron experiences in the presence of a metal wedge with $\Omega=3\pi/2$, and (c) $V_{\rm I}$, the attractive, `image-force potential' exerted by the metal wedge due to the presence of the electron. }
    \label{fig:V_Graphs}
\end{figure*}

Figure~\ref{fig:V_Graphs} demonstrates the potentials we can distinguish for an electron located at $(r_0,\theta_0,z_0)$. We refer to the potential due only to the original electron as the `Coulomb potential'  $V_{\rm C}$, the potential in the presence of the metal as the `potential' $V$, and the potential resulting from only the charge induced on the metal as the `image-force potential' $V_{\rm I}$, as illustrated in Fig.~\ref{fig:V_Graphs}a, b, and c, respectively. The three potentials are related using $V_{\rm I} = V-V_{\rm C}$ and are a function of position $(r,\theta,z)$ in addition to the location of the electron, $(r_0,\theta_0,z_0)$. To prevent the derivation from becoming unwieldy, we do not always explicitly mention $r,\theta,z;r_0,\theta_0,z_0$ as arguments for $V$, $V_{\rm I}$, or $V_{\rm C}$.

IFBL potential energy has historically been computed as $U_{\rm IFBL} = e\int_{\infty}^{x} {\mathcal E}_x  {\rm d}x_0$, where ${\mathcal E}_x = -{\rm d}V_{\rm I}/{\rm d}x$ is the electric field in the $x$-direction due to the image charge.\cite{sze2021physics} For our purpose, the calculation using ${\mathcal E}_x$ is cumbersome, even when taking the electric field in the $r$-direction instead of the $x$-direction, since it requires the evaluation of an additional derivative and integral. Instead, we obtain the result without performing the additional derivative and integral using
\begin{equation}
    U_{\rm IFBL}(r,\theta) = -\int_{0}^{-e} \frac{q}{e} V_{\rm I}(r,\theta,z;r,\theta,z){\rm d}q = -\frac{1}{2}eV_{\rm I}
\label{eq:VIF}
\end{equation}
where instead of bringing the electron from infinity to $(r,\theta,z)$, we increase the charge $q$ from $0$ to $-e$. As the charge is increased, the IFBL potential energy increases proportionally, which is captured by the prefactor $-q/e$. Since there is translational symmetry along the $z$-direction, the integral is independent of $z$.

Assuming a homogeneous dielectric environment, the potential $V(r)$ satisfies the Poisson equation $\nabla^2{V}=-\rho/\epsilon$, where the charge density $\rho$ is a point charge. 
Using separation of variables and employing cylindrical coordinates, the homogeneous solutions are $V_{\rm hom}(r,\theta,z)=R(r)\Theta(\theta)Z(z)$ and satisfy
\begin{equation}
\frac{\nabla^2 V_{\rm hom}}{V_{\rm hom}} = \frac{1}{R} \frac{\partial^2 R}{\partial r^2} + \frac{1}{rR}\frac{\partial R}{\partial r} + \frac{1}{r^2\Theta}\frac{\partial^2\Theta}{\partial \theta^2}  + \frac{1}{Z} \frac{\partial^2Z}{\partial z^2} = 0.
\end{equation}
Separating the $z$-dependent term yields the ordinary differential equation (ODE) ${\rm d}^2Z/{\rm d} z^2 = -k_z^2Z$, introducing the parameter $k_z$. Requiring the solution to be even around $z=z_0$ yields $Z_{k_z}(z)=\cos(k_z (z-z_0))$. Substituting $ Z_{k_z}$ back into the Laplacian and multiplying with $r^2$ separates the second ODE ${\rm d}^2\Theta/{\rm d} \theta^2  = \alpha^2\Theta$ and yields a third ODE $r^2 {\rm d}^2R/{\rm d} r^2 + r {\rm d} R/{\rm d} r - (k_z^2 r^2 + (i \alpha)^2) R = 0$, revealing that $R$ is a modified Bessel function of the first or second kind with imaginary order, {\it i.e.} $I_{i\alpha}(k_z r)$ or $K_{i\alpha}(k_z r)$. Since $ \lim_{r\rightarrow \infty} I_{i\alpha}(k_z r) \rightarrow \infty$, modified Bessel functions of the first kind can be discarded and $R_{\alpha k_z}(r) = K_{i\alpha}(k_z r)$.

A general solution of $V(r,\theta,z)$ is found by introducing the prefactor $e/\epsilon$, multiplying with a coefficient $C_{\alpha, k_z}$, and integrating over $\alpha$ and $k_z$
\begin{equation}
    V(r, \theta, z) = \frac{e}{\epsilon}\int_{0}^{\infty} {\rm d}k_z Z_{k_z}(z) \int_0^\infty {\rm d}\alpha C_{\alpha, k_z} \Theta_\alpha(\theta) R_{\alpha k_z}(r).
\label{eq:V1Final}
\end{equation}
After substituting back into the Poisson equation with the point charge located at ${\bf r}_0$ and using $ r^2{\rm d}^2 R/{\rm d} r^2+r{\rm d} R/{\rm d} r - k^2_zr^2R=-\alpha^2 R$ we obtain
\begin{multline}
    \int_{0}^{\infty} {\rm d}k_z Z_{k_z}(z) \int_{0}^{\infty}  {\rm d}\alpha C_{\alpha,k_z} R_{\alpha k_z}(r) \left(\frac{{\rm d}^2}{{\rm d} \theta^2} - \alpha^2\right) \Theta_\alpha(\theta) \\
    = r \delta(r - r_0) \delta(\theta - \theta_0) \delta(z - z_0).
    \label{eq:big_eq}
\end{multline}

To proceed, we must rewrite the right-hand side in the same basis as $V$ in Eq.~(\ref{eq:V1Final}). This is done using the Fourier cosine transform, ${\delta(z-z_0)} = \pi^{-1} \int_{0}^{\infty} {\cos( k_z (z-z_0))} {\rm d}k_z$, and the Kontorovich-Lebedev transform,  ${r\delta(r-r_0)}=2\pi^{-2}\int_{0}^{\infty} K_{i\alpha}(k_z r) K_{i\alpha}(k_z r_0) \sinh{(\pi \alpha)} \alpha {\rm d}\alpha$.\cite{kontorovich1938one} Because we have applied transforms, the coefficients on the left and right must be equal for each $\alpha$ and $k_z$. We find $C_{\alpha ,k_z}=2\pi^{-3}K_{i\alpha}(k_zr_0)\sinh(\pi\alpha)\alpha$ and are left with the ODE
\begin{equation}
    \left(\frac{{\rm d}^2}{{\rm d} \theta^2} - \alpha^2 \right) \Theta_{\alpha}(\theta) = \delta(\theta-\theta_0).
\label{eq:diffTheta}
\end{equation}

The boundary conditions for a point charge in free space are that $\Theta_{{\rm C};\alpha}(\theta)$ must be continuous everywhere, including $\theta=\theta_0$ and $\theta=\theta_0\pm\pi$. Additionally, the first derivative must be continuous except for $\theta=\theta_0$, where the Dirac delta introduces the discontinuity $\frac{\partial}{\partial\Theta}\Theta_\alpha\big|_{\theta=\theta_0+\eta}-\frac{\partial}{\partial\Theta}\Theta_\alpha\big|_{\theta=\theta_0-\eta}=1$ for $\eta\rightarrow0$.
Through inspection, we determine that ${\cosh(\alpha(\pi-|\theta-\theta_0|))}$ meets the continuity conditions. To meet the first derivative condition at $\theta=\theta_0$, we introduce a prefactor yielding 
\begin{equation}
    \Theta_{{\rm C} ; \alpha}(\theta;\theta_0)=-\frac{\cosh(\alpha(\pi-|\theta-\theta_0|))}{2\alpha\sinh(\alpha\pi)}
    \label{eq:homogenTheta}
\end{equation}
where the index C indicates that this is for the Coulomb kernel and no metal is considered. Substituting $\Theta_{\rm C}$ into Eq. (\ref{eq:V1Final}), we can verify the correctness of our solution since the Poisson kernel in cylindrical coordinates $-e/\left(4\pi\epsilon\sqrt{r^2+r_0^2+2rr_0\cos(\theta-\theta_0)+z^2}\right)$ is recovered. In essence, we have rewritten the Coulomb kernel in the form of Eq. (\ref{eq:V1Final}), where it can easily be modified to satisfy different boundary conditions at the metal surfaces.

When the metal plates are introduced, the boundary conditions are $V(r,0,z)=V(r,\Omega,z)=0$, which straightforwardly translate to $\Theta_\alpha(0)=\Theta_\alpha(\Omega)=0$, whereas the boundary conditions at $\theta_0=\theta$ are unchanged, ${\it i.e.}$ a continuity of $\Theta(\theta)$ and a discontinuity of the first derivative. We add two homogeneous solutions to Eq. (\ref{eq:homogenTheta}), $\sinh(\alpha\theta)$ and $\sinh(\alpha(\Omega-\theta))$, and scale with appropriate prefactors, which are chosen so that $\Theta_\alpha(\Omega)=0$ and $\Theta_\alpha(0)=0$, respectively. We restrict $\theta_0 \in [0,\Omega]$ and find $\Theta_\alpha(\theta;\theta_0)=\Theta_{{\rm C};\alpha}(\theta;\theta_0)+\Theta_{{\rm I};\alpha}(\theta;\theta_0)$ with

\begin{align}
    \Theta_{{\rm I};\alpha}(\theta;\theta_0)=
    \frac{\sinh(\alpha\theta)\cosh\big(\alpha(\pi-(\Omega-\theta_0))\big)}{2\alpha\sinh(\alpha\pi)\sinh(\alpha\Omega)}\nonumber\\
    + \frac{\sinh(\alpha(\Omega-\theta))\cosh\big(\alpha(\pi-\theta_0)\big)}{2\alpha\sinh(\alpha\pi)\sinh(\alpha\Omega)}.
\label{eq:FinalTheta}
\end{align}
Substituting $C_{\alpha,k_z}$ and Eq.~(\ref{eq:FinalTheta}) into Eq.~(\ref{eq:V1Final}) and taking the cosine Fourier transform of $K_{i\alpha}(k_zr)K_{i\alpha}(k_zr_0)$ yields
\begin{align}
    V_{\rm I}=\frac{e}{4\pi\epsilon\sqrt{rr_0}}\int_0^\infty {\rm d}\alpha P_{i\alpha-1/2}\left(\frac{(z-z_0)^2+r_0^2+r^2}{2rr_0}\right)\nonumber\\
    \bigg( \frac{\sinh(\alpha\theta)\cosh\big(\alpha(\pi-(\Omega-\theta_0))\big)}{\sinh(\alpha\Omega)\cosh(\alpha\pi)}\nonumber\\
    +\frac{\sinh(\alpha(\Omega-\theta))\cosh\big(\alpha(\pi-\theta_0)\big)}{\sinh(\alpha\Omega)\cosh(\alpha\pi)}\bigg)
\end{align}
where $P_{n}(z)$ the Legendre functions of the first kind.\cite{Bateman} Setting $r=r_0$, $z=z_0$, and $\theta=\theta_0$, the IFBL potential energy of an electron is
\begin{align}
    U_{\rm IFBL}&(r, \theta) = \frac{-e^2}{8\pi\epsilon r}\int^{\infty}_{0} {\rm d}\alpha \nonumber
    \bigg(\frac{\sinh(\alpha\theta) \cosh(\alpha(\pi-(\Omega-\theta)))}{\sinh(\alpha\Omega)\cosh(\alpha\pi)}
    \nonumber\\
    & +\frac{\sinh(\alpha(\Omega-\theta))\cosh(\alpha(\pi-\theta))}{\sinh(\alpha\Omega)\cosh(\alpha\pi)}\bigg) 
    \label{eq:FinalEquation}
\end{align}
since the argument of the Legendre function becomes 1, and $P_{i\alpha - 1/2}(1)=1$.

\subsection{Method of Images on Cone Manifold}
\begin{figure}
    \centering
    \includegraphics[width=8.6cm]{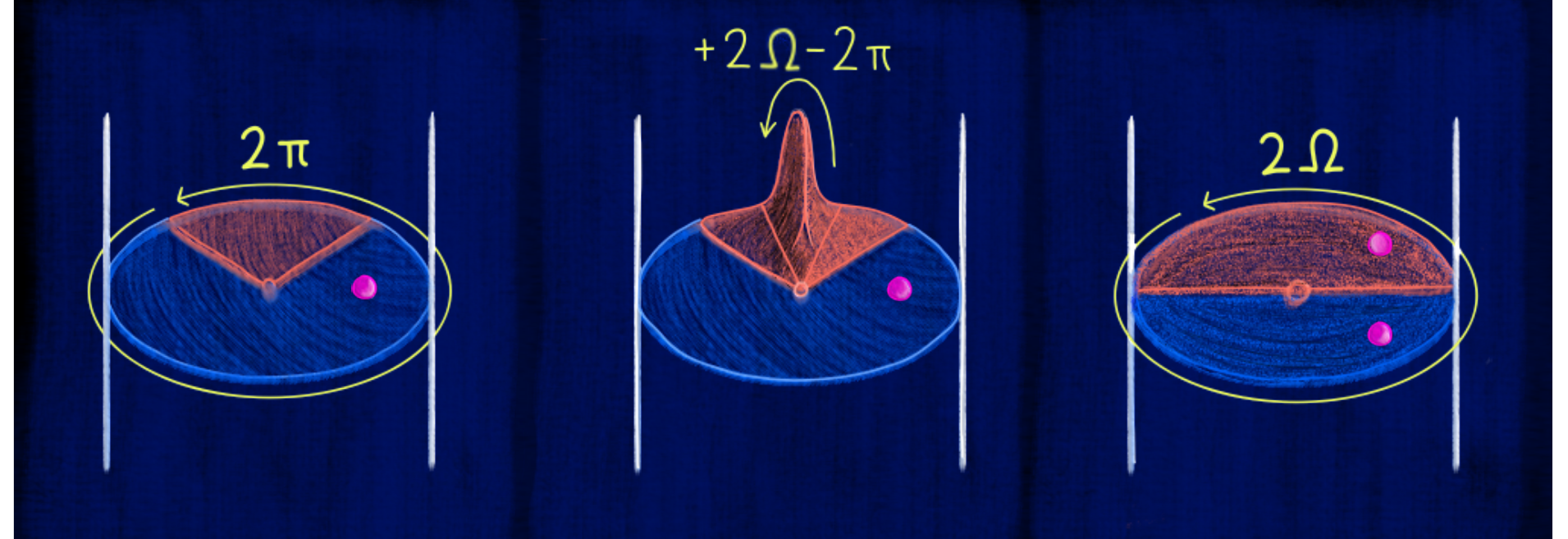}
    \caption{Illustration demonstrating, from left to right, how expanding Euclidean space with $2\Omega-2\pi$ in the metal (orange region) creates a cone manifold space, enabling image charge placement opposite an electron (pink dot) in a symmetrical position relative to the metal's boundary.}
    \label{fig:ConifoldTheta}
\end{figure}

We shall demonstrate next that the method of images can also be used with only a single image provided a non-Riemannian space is used. More precisely, a space requiring a total rotation of $2\Omega$ around the origin to arrive at the same location instead of $2\pi$, as illustrated in Fig.~\ref{fig:ConifoldTheta}. This non-Riemannian space is a generalization of an orbifold\cite{thurston1979geometry} and is known as a `cone manifold'.\cite{jones1990cone} Mapping the space in the dielectric, {\it i.e.} $\theta \in [0,\Omega]$,  onto one half of the cone manifold, the metal surface now appears perfectly flat, creating symmetry that allows for an image charge to be easily placed on the other side of the metal. We note that Sommerfeld also extended the method of images to a non-Euclidean, albeit still Riemannian space, where he used analytical continuation to determine the Coulomb kernel.\cite{von1896verzweigte,davis1971solution,alshal2021image} However, Sommerfeld's method requires a sum over $m$ terms where $\theta=m\pi/k$ and $m$ and $k$ are integers.

Repeating the derivation for the Coulomb kernel in the cone manifold and now requiring continuity at $\theta=\theta_0\pm\Omega$ gives
\begin{equation}
    \Theta^\circledast_{{\rm C} ; \alpha}(\theta;\theta_0)=-\frac{\cosh(\alpha(\Omega-|\theta-\theta_0|))}{2\alpha\sinh(\alpha\Omega)}
    \label{eq:homogenThetaprime}
\end{equation}
where the $\circledast$ indicates that this $\Theta^\circledast$ is defined on the cone manifold and $\theta\in[0,2\Omega]$. The image will have
\begin{equation}
    \Theta^\circledast_{{\rm im} ; \alpha}(\theta;\theta_0)=-\frac{\cosh(\alpha(\Omega-|-\theta-\theta_0|))}{2\alpha\sinh(\alpha\Omega)}.
    \label{eq:homogenThetaprime}
\end{equation}
The method of images now immediately yields the potential for the geometry with the metal by adding the same potential with an opposite sign at the other side of the metal plate, {\it i.e.} $V(r,\theta,z;r_0,\theta_0,z_0)=V_{\rm C}^\circledast(r,\theta,z;r_0,\theta_0,z_0)-V^\circledast_{\rm C}(r,\theta,z;r_0,-\theta_0,z_0)$. The way to determine the image-force potential remains $V_{\rm I}=V-V_{\rm C}$. Evaluating for $r=r_0,z=z_0$, and $\theta=\theta_0+\eta$ yields
\begin{align}
    U_{\rm IFBL}(r,\theta)=\frac{-e^2}{8\pi \epsilon r}\int_0^\infty {\rm d}\alpha\bigg( \frac{\cosh(\alpha(\Omega-\eta))\sinh(\alpha\pi)}{\sinh(\alpha\Omega)\cosh(\alpha\pi)} -\nonumber\\
    \frac{\cosh(\alpha(\Omega-2\theta))\sinh(\alpha\pi)}{\sinh(\alpha\Omega)\cosh(\alpha\pi)} - \frac{\cosh(\alpha(\pi-\eta))}{\cosh(\alpha\pi)}\bigg)
\end{align}
where we introduced $\eta\to 0$ to avoid the singularity that appears before the first and last terms are subtracted. After the subtraction, we obtain
\begin{multline}
    U_{\rm IFBL}(r,\theta) = \frac{-e^2}{8\pi \epsilon r} \bigg(\int^\infty_0 {\rm d}\alpha \frac{\sinh(\alpha(\Omega-\pi))}{\sinh(\alpha\Omega)\cosh(\alpha\pi)} \\
     +\int^\infty_0 {\rm d}\alpha \frac{\cosh(\alpha(\Omega-2\theta))\tanh(\alpha\pi)}{\sinh(\alpha\Omega)}\bigg) \label{eq:finaleq2}
\end{multline}
which is to equal the IFBL potential energy in Eq.~(\ref{eq:FinalEquation}) through trigonometric identities.

\subsection{Numerical Implementation}
The first integral in Eq.~(\ref{eq:finaleq2}) is independent of $\theta$ and can be evaluated in closed form for some $\Omega$. Obviously for $\Omega=\pi$, the first integral vanishes but for $\Omega=3\pi/2$ the integral evaluates to $1/2-2\sqrt{3}/9\approx 0.115$.

For arbitrary $\theta$ we evaluate the second integral of Eq.~(\ref{eq:finaleq2}) numerically. We observe that the asymptotic behavior of the integrand is $\alpha \to \infty$ is $e^{-2\alpha\theta}$. After  transforming $t = 2\alpha\theta$, the integral  becomes
\begin{equation}
    \int^\infty_0 {\rm d}\alpha \frac{\cosh(\alpha(\Omega-2\theta))\tanh(\alpha\pi)}{\sinh(\alpha\Omega)} = \int_0^{\infty} {\rm d} t f(t) e^{-t} \label{eq:part2}
\end{equation}
where
\begin{equation}
f(t) = \frac{\tanh(\pi t/(2\theta))(1+{\rm e}^{-t(\Omega/\theta-2)})}{2\theta(1-{\rm e}^{-t\Omega/\theta })}.
\end{equation}
The integral in Eq.~(\ref{eq:part2}) can be evaluated numerically as $\sum_i w_i f(t_i)$ where $t_i$ are the $i$-th roots of the Laguerre polynomial, $w_i$ are the weights,\cite{salzer1949table} and we choose $N=15$. We go from Cartesian to polar coordinates using $r=\sqrt{x^2+y^2}$ and $\theta=\arg{(x+{\rm i}y)}$. For $\theta>\Omega/2$ we use $U_{\rm IFBL}(r,\theta)=U_{\rm IFBL}(r,\Omega-\theta)$.


\begin{figure}[t]
    \centering
    \includegraphics[width=8.6cm]{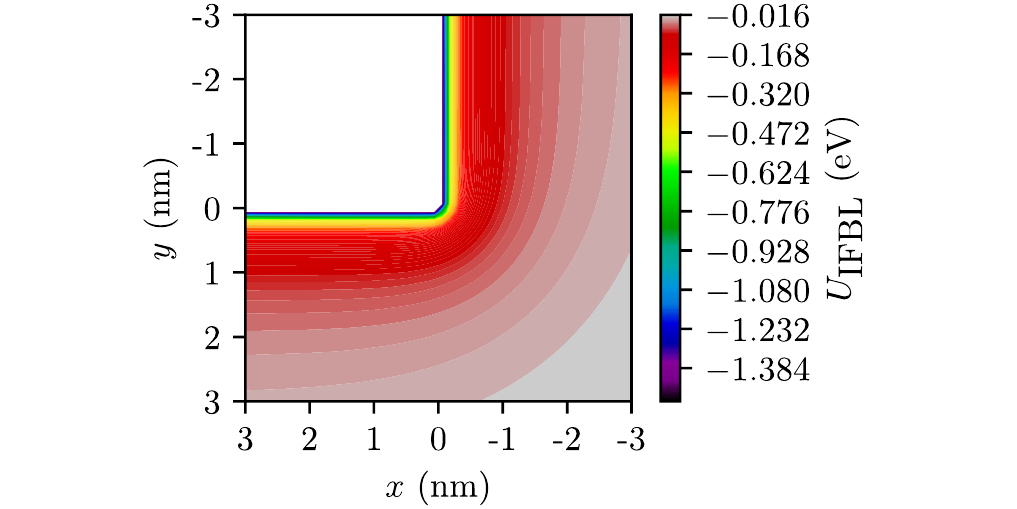}
    \caption{Graph of the IFBL potential energy of an electron as a function of $x$ and $y$ with $\Omega=3\pi/2$ and $\epsilon=3.9\epsilon_0$, which is the permittivity of SiO$_2$.}        
    \label{fig:Image_Potential}
\end{figure}

\begin{table}
\begin{tabular}{c|c|c|c}
\hline
     $\ $ & $\Omega$ & $\theta$ & $U_{\rm IFBL}/(-e^2/(16\pi\epsilon r)) $ \\
     \hline
     $a$ & $\pi$ & $\pi/2$ & 1 \\
     $b$ & $\pi$ & $\theta$ & $\csc(\theta)$\\
     $c$ & $2\pi$ & $\pi$ & $2/\pi\approx 0.63$\\
     $d$ & $3\pi/2$ & $\pi$ (or $\pi/2$) & $6-2/\sqrt{3}\approx 0.85$ \\
     $e$ & $4\pi/3$ & $\pi$ & $8\sqrt{3}/9+1/\pi-3/4 \approx 1.11$ \\
     $f$ & $2\pi$ & $3\pi/2$ & $1/\pi+1/2 \approx 0.82$ \\
     $g$ & $3\pi/2$ & $\Omega/2=3\pi/4$ & $2+2\sqrt{2}-16\sqrt{3}/9\approx 0.74$\\
     $h$ & $4\pi/3$ & $\Omega/2=2\pi/3$ & $4/\sqrt{3}-3/2 \approx 0.81$ \\
     $i$ & $2\pi/3$ & $\Omega/2=\pi/3$ & $4\sqrt{3}/9+2/\pi \approx 1.41$ \\
     $j$ & $\pi/2$ & $\Omega/2=\pi/4$ & $2\sqrt{2}-1 \approx 1.83$\\
     $k$ & $\pi/3$ & $\Omega/2=\pi/6$ & $5-4/\sqrt{3} \approx 2.69$ \\
     \hline
\end{tabular}
\caption{Image-force barrier lowering for various angles between the metal surface, $\Omega$, and for various angles between the metal and the 2D material, $\theta$.}\label{tab:table_theta}
\end{table}
\begin{figure}[hbt!]
    \centering
    \includegraphics[width=8.6cm]{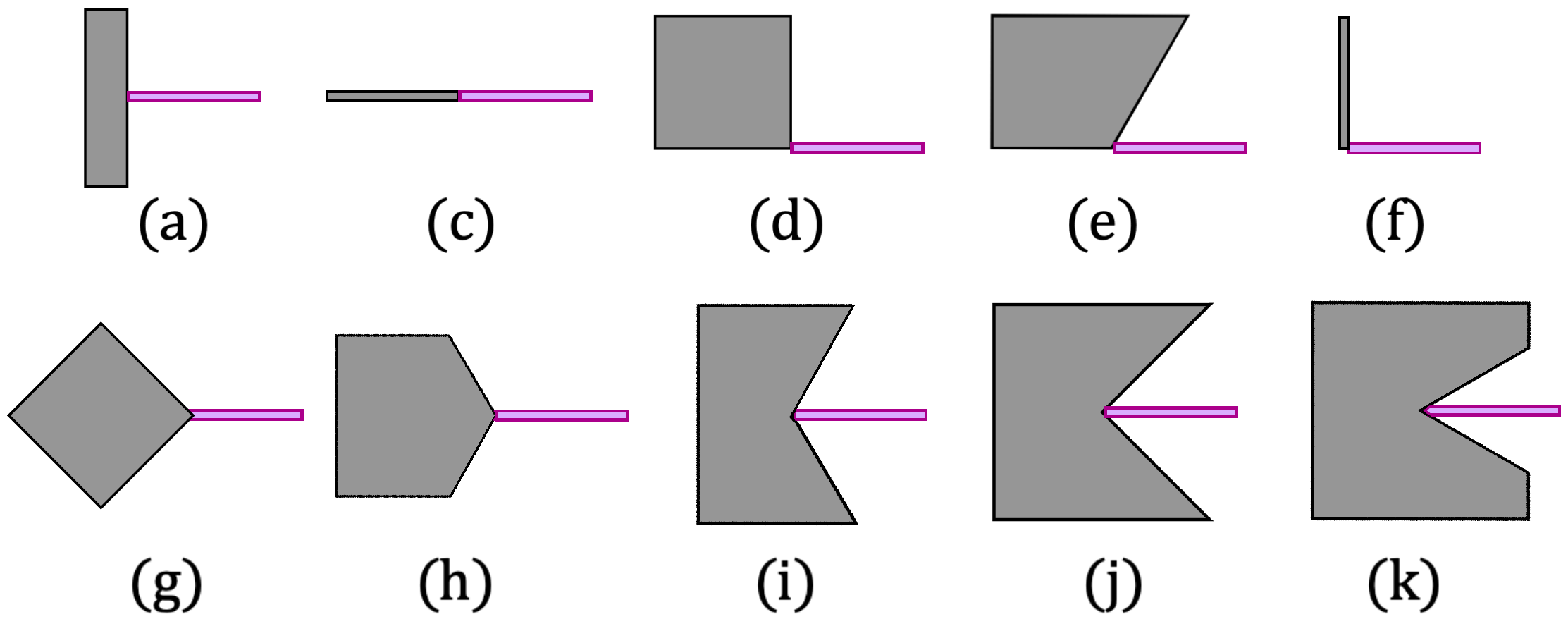}
    \caption{Illustrations depicting 10 different configurations considered in Table~\ref{tab:table_theta}. The metal surface (grey) is separated by an angle $\Omega$ and the 2D semiconductor (pink) has an angle $\theta$ with the upper-most metal surface. Illustrations are rotated such that the semiconductor is horizontal.}
    \label{fig:TableDrawings}
\end{figure}

\section{Results and Simulation}
Figure~\ref{fig:Image_Potential} is the plot of Eq.~(\ref{eq:finaleq2}) for $\Omega=3\pi/2$ assuming SiO$_2$ as the dielectric with a relative permittivity of 3.9. The metal is located in the upper left corner region where $x>0{\rm nm}$ and $y<0{\rm nm}$. 
Far removed from the corner, the traditional IFBL potential energy $-e^2/(16\pi\epsilon x)$ or $-e^2/(16\pi\epsilon y)$ is recovered but at the corner, the IFBL effect is reduced. Visual inspection shows that the barrier can easily be lowered by more than 0.1~eV due to the IFBL effect, which could improve contact resistance by orders of magnitude.

\subsection{IFBL for Different Contact Angles}
We evaluate some special cases of angles between the metal and the 2D semiconductor in Table~\ref{tab:table_theta}. Fig.~\ref{fig:TableDrawings} illustrates all the geometries listed in the table, except for b, where $\theta$ is variable. Taking $\Omega=\pi$ and $\theta=\pi/2$ recovers the usual IFBL potential energy expression $U_{\rm IFBL}^{\rm (a)}=-e^2/(16\pi r)$. Taking $\Omega=3\pi/2$ and $\theta=\pi$, as would be the case in Fig.~\ref{fig:TopAndSideContacts}b, only reduces the IFBL potential energy by 15\%. Choosing the 2D material to be in the same plane as one of the metal plates, {\it i.e} $\theta=\pi$ as is the case in Fig.~\ref{fig:TableDrawings}c-e, strengthens the IFBL potential energy relative to the Fig.~\ref{fig:TableDrawings}a case once $\Omega<1.384\pi$. And for $\Omega=4\pi/3$ or Fig.~\ref{fig:TableDrawings}e, {\it i.e.} when there is an angle $\pi/3$ between the metal and the 2D material, the IFBL potential energy is roughly 11\% stronger. The configuration in Fig.~\ref{fig:TableDrawings}k, which has the 2D material in between two plates with an angle $\pi/3$, improves the IFBL potential energy by a factor 2.69, which is the highest of all cases we consider.

\begin{figure*}
    \centering
    \includegraphics[width=\linewidth]{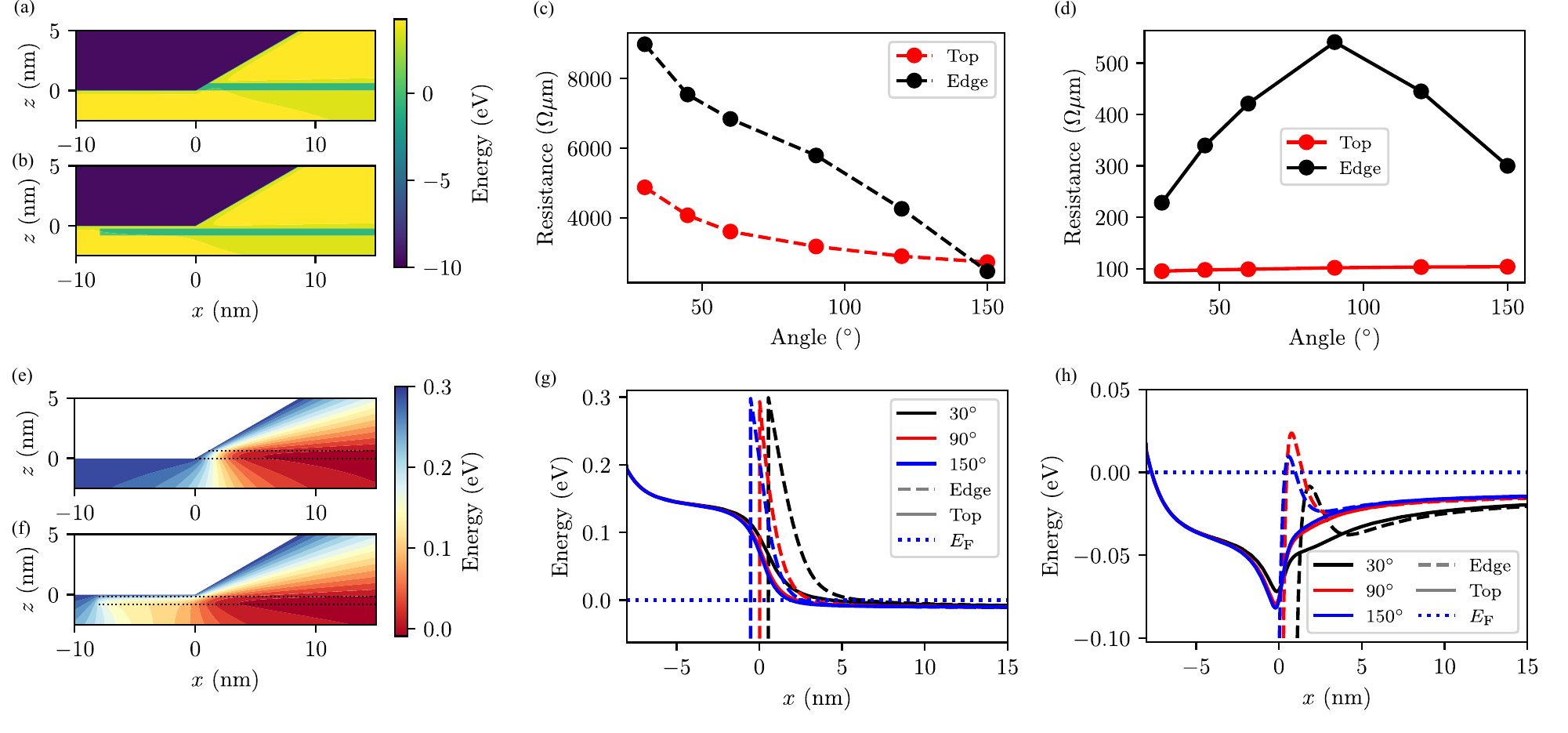}
    \caption{Potential energy landscape of (a) an edge and (b) a top contact with a $30 \degree$ angle between the upper metal plate and semiconductor in contact resistance calculations. Contact resistance is shown as a function of the angle between metal and semiconductor in a top and edge configuration (c) without considering the IFBL and (d) including the IFBL. Contour plot of numerical solution to Poisson equation in (e) edge and (f) top contact where the position the TMD is marked by the black-dotted lines. Potential energy profiles in the middle of the TMD as function of the $x$-direction (g) without IFBL and (h) with IFBL. Note that experimentally, IFBL will always be present and the contact resistance in (d) will most accurately describe experimental contact resistance.}
    \label{fig:ResistanceVAngle}
\end{figure*}

\subsection{Impact of IFBL on Contact Resistance}

Above, we focused solely on the IFBL. However,
the practical question is how contact resistance is influenced
by various semiconductor-metal configurations. It is important to note that in addition to IFBL, contact resistance is also affected by the potential profile. Higher electric fields at the interface give rise to a larger impact of the IFBL to determine the Schottky barrier. The simplified model is $\Delta \phi = \sqrt{\frac{e {\mathcal E}}{4\pi \epsilon}}$ where ${\mathcal E}$ is the electric field at the surface. However, in 2D materials, the depletion approximation can not be used and a more detailed calculation is required to obtain contact resistance.

To determine the potential profile in the 2D material, we solve the Poisson equation
\begin{equation}
    \nabla\cdot (\epsilon({\bf r}) \cdot \nabla V_{\rm contact}({\bf r})) = e (N_{\rm D}({\bf r}) - n({\bf r})) \label{eq:poiss_contact}
\end{equation}
self-consistently with the charge in the 2D material. Here, $ n({\bf r}) =  n_{\rm 2D} \vartheta({\bf r})/ t_{\rm TMD} $
where $\vartheta({\bf r})$ equals unity inside the TMD and vanishes outside of the 2D material, $t_{\rm TMD}$ is the thickness of the TMD, and
\begin{equation}
n_{\rm 2D} = \frac{2m^* k_\textrm{B} T}{\pi \hslash^2} \ln{\left[1+\exp{\left(\frac{E_\textrm{F} - V({\bf r})}{k_\textrm{B} T} \right)}\right]}
\end{equation}
where $m^*$ is the TMD effective mass, $T$ is the temperature, $k_{\rm B}$ the Boltzmann constant, $\hslash$ the reduced Planck constant, and $E_{\rm F}$ the Fermi level.  The factor of two arises due to the valley degeneracy of the K and K’ valleys in the first Brillouin zone of the MoS$_2$ lattice.

We solve the Poisson equation numerically in a 2D slice of the contact configuration using the finite element solver (FEM) \cite{LoggWells2010, LoggEtal_10_2012, AlnaesEtal2014} of the FEniCSx software package. 
We mesh the simulation domain using a finite element mesh generator Gmsh\cite{Gmsh} taking care that the meshing size is small enough compared to the smallest dimensions within the simulation domain (10 times smaller than the vdW gap).
The domain size is $150 \times 100$ nm, with Dirichlet boundary conditions on the metal surfaces and Neumann boundary conditions on the rest of the simulation boundary.
The Dirichlet boundary conditions of the metal are set to a Schottky barrier height of 0.3 eV.
The Fermi level is kept as a reference to 0.0 eV while we dope the TMD with a doping concentration of $N_\textrm{D} \cdot t_{\rm TMD} = 1 \times 10^{13} \ \rm cm^{-2}$.

We assume SiO$_2$ as the dielectric surrounding the 2D material with $\epsilon = 3.9 \epsilon_0$, where $\epsilon_0$ is the permittivity of free space. 
The 2D semiconductor material is assumed to be a monolayer of MoS$_2$ with $t_{\rm TMD} = 0.61$ nm \cite{Laturia2018} and effective mass of $0.5 m_e$ \cite{effectivemass1, effectivemass2} where $m_e$ is the electron mass. The dielectric constant tensor of MoS$_2$ has an in-plane component of $\epsilon_\parallel = 15.5 \epsilon_0$ and an out-of-plane component of $\epsilon_\perp = 6.2 \epsilon_0$.\cite{Laturia2018} In the FEM solver, we use the dielectric constant of $\epsilon = 3.9 \epsilon_0$ in the SiO$_2$ and the tensor in the MoS$_2$. For IFBL, the derivation does not admit inhomogeneous dielectrics; we use the dielectric constant of a SiO$_2$ environment ($3.9 \epsilon_0$) throughout, which was shown to be a reasonable approximation.\cite{Brahma2023}

In the top contact configuration, we include the presence of a van-der-Waals (vdW) gap with a thickness of $0.2$ nm and a barrier height of 4 eV, while the metal/semiconductor overlap region is 8 nm long.


We calculate the contact resistance $\rho_{\rm c}$ according to
\begin{equation}
    \frac{1}{\rho_{\rm c}} = \frac{2e^2}{h} \int_{-\infty}^{\infty} dE \left| \frac{df(E)}{dE} \right|  \int_{-\infty}^{\infty} \frac{dk_y}{2\pi} T(k_y, E)
\end{equation}
where $f(E)$ is the Fermi-Dirac distribution and $T(k_y, E)$ is the transmission obtained from the quantum transport calculations at a particular energy $E$ and wavenumber in the $y$-direction $k_y$.\cite{Brahma2023} We integrate the transmission over an energy window of $-20 k_{\rm B}T$ to $20 k_{\rm B} T$ at energy steps of $0.315 k_{\rm B} T$ and $k_y$ over an equivalent energy window.

The transport calculation is based on the quantum boundary transmitting method (QTBM) \cite{QTBM, QTBMoriginal, VANDEPUT2019156} which is equivalent to the ballistic non-equilibrium Green's function (NEGF) method.
We use an effective mass description of the Hamiltonian which will be represented by a sparse matrix of size $N_xN_z \times N_xN_z$. Using the QTBM, we add infinite contact leads to the left, right, and top boundaries of our system. We then inject modes into the right contact lead through the 2D material. At small enough injection energies E, only one traveling mode exists within the TMD quantum well due to energy confinement. Thus, we only need to inject one mode, $B_{\rm in}$, per energy step. We solve for the wavefunction $c$ throughout the device as
\begin{equation}
    [EI-H-\Sigma]c = B
    \label{eq:InjWave}
\end{equation}
where $E$ is the energy of the injected mode, $I$ is the identity matrix, $H$ is the Hamiltonian of the full contact region, and $\Sigma$ are the contact self-energies at the boundary of the simulation region. $B$ is an injection term $B = [0_{N_x\times (N_{z}-1)}, B_{\rm in}]^T$ where we only have a single mode injected from the right. Because only one mode is injected per energy, we only have to solve Eq.~(\ref{eq:InjWave}) once.


To analyze the impact of IFBL, we compute current both with and without IFBL. The potential energy is the potential determined from Eq.~(\ref{eq:poiss_contact}) and when IFBL is considered, the IFBL potential energy from Eq.~(\ref{eq:finaleq2}) is added: $U_{\rm QTBM}=-eV_{\rm contact}({\bf r})+U_{\rm IFBL}({\bf r})$. The simulation domain is decreased after solving the Poisson equation in the interest of computational performance to $25 \times 7.5$ nm$^2$ with rectangular mesh spacings of $\Delta x = 0.025$ nm and $\Delta z = 0.0125$ nm.

Figures \ref{fig:ResistanceVAngle}a and b show the potential energy landscape of the edge and top contact, respectively.
The potential energy landscape of the transport domain is given by the sum of the previously calculated solutions of the Poisson equation shown in Fig. \ref{fig:ResistanceVAngle}e-f, the IFBL, and the conduction band minima of the materials.
Keeping the Fermi level as reference at 0.0 eV, we assume that the metal work function is 4.30 eV while the conduction band minimum is set to -10 eV. For the oxide, we assume that the conduction band minimum is set to 4.0 eV while the electron affinity of the MoS$_2$ monolayer is set to 4.0 eV, leading to a Schottky barrier height of 0.3 eV according to the Schottky-Mott rule.

Figure \ref{fig:ResistanceVAngle}c shows the contact resistance as a function of the angle between the upper metal plate and the MoS$_2$ monolayer without accounting for IFBL.
The contact resistance in the top and edge contact decreases as the angle between the metal and the semiconductor increases.
At angles around $150\degree$, the top and edge contact exhibit similar contact resistances, whereas at smaller angles, the contact resistance is lower in the top contact compared to the edge contact. Note that experimentally IFBL will always be present experimentally and that the results in Fig.~\ref{fig:ResistanceVAngle}c are presented here to quantify the impact of IFBL.

Figure \ref{fig:ResistanceVAngle}g shows the potential energy profile in the $x$-direction in the middle of the TMD for contact configurations with different angles between the metal and TMD without IFBL.
The dashed and full lines show the potential profiles for an edge- and top-contacted TMD, respectively.
For both the edge and top contacted TMD, the depletion length decreases as the angle increases due to the metal plate and by extension, the Schottky barrier being further away from the TMD channel.
Comparing the effective Schottky barrier height in the TMD for the edge and top contact, we conclude that the Schottky barrier height is lower for the top contact by about $\sim 0.15$ eV due to the top contact being further separated from the metal by the vdW gap.

Figure \ref{fig:ResistanceVAngle}d demonstrates the contact resistance vs. the angle between the metal and the MoS$_2$ monolayer when the IFBL is added to the energy landscape as is shown in Fig. \ref{fig:ResistanceVAngle}a and b.
The impact of adding IFBL significantly lowers contact resistance by an order of magnitude for both edge contacts and top contacts, respectively. We find that top contacts exhibit relatively constant contact resistance regardless of angle. Edge contacts experience lower contact resistance at angles smaller and, interestingly, larger than 90$\degree$.

Figure \ref{fig:ResistanceVAngle}h shows the potential energy profile in the $x$-direction in the middle of the TMD for different angles when the IFBL is added. For the top contacts, underneath the metal ($x<0~$nm), the IFBL is constant and the energy profile as shown in Fig.~\ref{fig:ResistanceVAngle}g is shifted downwards in Fig.~\ref{fig:ResistanceVAngle}h. Past $x>0~$nm, IFBL starts to decrease towards the right hand side where the potential is unaffected by the IFBL and the potential energy in Fig.~\ref{fig:ResistanceVAngle}g and Fig.~\ref{fig:ResistanceVAngle}h are similar.

Compared to the potential energy profiles demonstrated in Fig. \ref{fig:ResistanceVAngle}g, the effective Schottky barrier is significantly lowered by about $\sim 0.30$ eV and $\sim 0.25$ eV for edge and top contacts, respectively, leading to lower contact resistances as seen Fig. \ref{fig:ResistanceVAngle}d. The edge contact profiles illustrate that at large and small angles the effective Schottky barrier height is lower compared to the $90 \degree$ case, resulting in the contact resistance behavior shown in Fig.~\ref{fig:ResistanceVAngle}d. The top contact energy profiles show an absence of an energy barrier in the $x$-direction of the TMD, leading to a nearly constant contact resistance as a function of the metal plate angle. In contrast to what one would expect when only considering the depletion potential, contact resistance does not increase with decreasing angle since the IFBL counteracts the increase in the depletion length within the TMD.

\section{Discussion}

Interestingly, top contacts have lower contact resistance compared to edge contacts both with and without IFBL. For the top contact, the doping is underneath the entire bottom of the metal and the voltage drop over the vacuum between the 2D material and the metal is significant, leaving a much lower Schottky barrier in the 2D material. For the edge contact, the metal is in contact with the (undoped) dielectric; there is no voltage drop and the 2D material sees the full Schottky barrier.

Previously, we showed that smaller angles between the metal and semiconductor would strengthen IFBL significantly. 
However, due to the impact of the electrostatics on contact resistance, we find that this increase in IFBL does not always translate to a decrease in contact resistance. 
In our simulation, for edge contacts, both small and large angles decrease the contact resistance. For large angles, the decrease in depletion barrier is more important than the weakening of the IFBL, whereas at small angles the strengthening of the IFBL outweighs the impact of the longer depletion layer. 
For edge contacts, both smaller and larger angles result in smaller contact resistance.

Comparison of theoretical predictions for contact resistance to experiments remains very challenging because many factors have a large impact on contact resistance. The most important parameters we identify are: contact metal workfunction, electron affinity of the 2D material, doping concentration, permittivity of the surrounding material, and geometry. Note that we assume a contact to a gapped material whereas the physics of making contacts to graphene\cite{wang2013one} will be significantly different. Independently characterizing all important parameters experimentally will be very challenging. The recent approach of Vaknin {\it et al.}, determining the potential profile\cite{vaknin2020schottky} in Au-MoS$_2$ contacts using Kelvin probe force microscopy, may be a way to compare more directly to a simulated potential profile. Trying to completely eliminate the Schottky barrier is one strategy to lower contact resistance\cite{shen2021ultralow,li2023approaching} but the use of lower workfunction metals may be problematic in terms of materials stability of a practical technology.


\section{Conclusions}
In summary, because 2D material contacts are strongly affected by the permittivity of the surrounding material, IFBL must be incorporated to accurately model the Schottky barrier height. We determined the IFBL potential energy emerging from a metal with surfaces separated by an angle $\Omega$. We presented two methods for determining the IFBL potential energy, one using a direct determination and the other using the method of images provided a cone-manifold space is used. To facilitate practical implementation, we provided a numerical evaluation of Eq.~(\ref{eq:finaleq2}) using Gauss-Laguerre quadrature. We then considered 10 configurations with various angles between the metal surfaces and the 2D material where the IFBL potential energy can be evaluated analytically. We found that the IFBL potential energy was weakened by a factor $6-2\sqrt{3}\approx 0.85$ for a top metal contact with its surface perpendicular to the 2D material and found that IFBL potential energy is strengthened by 11\% for a metal contacting the semiconductor with an angle $\pi/3$ between them.

We also calculated the contact resistance experienced by various geometric configurations. We found that for edge contacts, a non-perpendicular metal contact angle yielded lower contact resistance, however, top contacts experienced lower contact resistance than edge contacts for all angles. We note that a stronger IFBL barrier does not always translate to a decrease in contact resistance because the depletion region is increased by the smaller contact angle. We identified contact angle as a tunable parameter that can impact IFBL energy and therefore the contact resistance of 2D semiconductor-metal contacts. In all cases, the strength of the IFBL potential energy is also scaled with the permittivity of the material surrounding the 2D material, highlighting the importance of using a dielectric with low permittivity surrounding the 2D material to minimize the Schottky barrier. 

\section{Acknowledgements}
This material is based upon work supported by the Taiwan Semiconductor Manufacturing Company, Ltd.

\bibliography{citations2.bib}
\bibliographystyle{ieeetr}

\end{document}